\documentclass{article}
\usepackage{emulateapj}
\usepackage{amsmath}
\usepackage{amssym}
\usepackage{psfig}
\usepackage{times}

\input{epsf}

\def\Cp{C_{\rm peak}}

\newbox\grsign \setbox\grsign=\hbox{$>$} \newdimen\grdimen \grdimen=\ht\grsign
\newbox\simlessbox \newbox\simgreatbox \newbox\simpropbox
\setbox\simgreatbox=\hbox{\raise.5ex\hbox{$>$}\llap
     {\lower.5ex\hbox{$\sim$}}}\ht1=\grdimen\dp1=0pt
\setbox\simlessbox=\hbox{\raise.5ex\hbox{$<$}\llap
     {\lower.5ex\hbox{$\sim$}}}\ht2=\grdimen\dp2=0pt
\setbox\simpropbox=\hbox{\raise.5ex\hbox{$\propto$}\llap
     {\lower.5ex\hbox{$\sim$}}}\ht2=\grdimen\dp2=0pt

\def\simlt{\mathrel{\copy\simlessbox}}

\lefthead{BELOBORODOV, STERN, \& SVENSSON}
\righthead{POWER DENSITY SPECTRA OF GAMMA-RAY BURSTS}

\begin{document}

\title{Power Density Spectra of Gamma-Ray Bursts}

\author{Andrei M. Beloborodov\altaffilmark{1}, Boris E. Stern\altaffilmark{2},
and Roland Svensson\altaffilmark{3}}
\altaffiltext{1}{Also at the Astro-Space Center of Lebedev Physical Institute,
Profsoyuznaya 84/32, Moscow 117810, Russia}
\altaffiltext{2}{Institute for Nuclear Research of Russian Academy of Science,
Profsojuznaja 7a, 117312 Moscow, Russia}
\altaffiltext{3}{Institute of Theoretical Physics, UCSB,
Santa Barbara CA 93106, USA}

\affil{Stockholm Observatory, SE-133 36, Saltsj{\"o}baden, Sweden}

\begin{abstract}
Power density spectra (PDSs) of long gamma-ray bursts (GRBs)
provide useful information on GRBs, indicating their self-similar temporal
structure.  The best power-law PDSs are displayed by the longest
bursts ($T_{90}>100$~s) in which the range of self-similar time scales covers
more than  2 decades. Shorter bursts have apparent
PDS slopes more strongly affected by statistical fluctuations.
The underlying power law can then be reproduced with high accuracy
by averaging the PDSs for a large sample of bursts.
This power law has a slope $\alpha\approx-5/3$ and a sharp break at
$\sim 1$~Hz.

The power-law PDS provides a new sensitive tool for studies of GRBs.
In particular, we calculate the PDSs of bright bursts in separate
energy channels.
The PDS flattens in the hard channel ($h\nu>300$ keV) and steepens in the
soft channel ($h\nu<50$ keV), while the PDS of bolometric light curves
approximately follows the $-5/3$ law.

We then study dim bursts and compare them to the bright
ones. We find a strong correlation between the burst brightness and the PDS
slope. This correlation shows that the bursts are far from being standard
candles and dim bursts should be {\it intrinsically} weak.
The time dilation of dim bursts is probably related to physical
processes occurring in the burst rather than to a cosmological redshift.
\end{abstract}


\section{Introduction}

The light curves of gamma-ray bursts (GRBs) typically have many random peaks
and in spite of extensive statistical studies (e.g., Nemiroff et al. 1994;
Norris et al. 1996; Stern 1996), the temporal behavior of GRBs remains a
puzzle.
Contrary to the complicated diverse behavior in the time domain, long GRBs show
a simple behavior in the Fourier domain (Beloborodov, Stern, \& Svensson 1998).
Their PDS is a power law of index $\alpha\approx -5/3$ (with a break at
$\sim 1$~Hz) plus standard (exponentially distributed) statistical fluctuations
superimposed onto the power law. The PDS slope and the break
characterize the process randomly generating the diverse light curves of GRBs.
Intriguingly, the PDS slope coincides with the slope of the Kolmogorov law.
The power-law behavior is seen in individual bursts (we illustrate
this in \S 3) and may provide a clue to the nature of GRBs.

In the present paper, we study in detail the PDSs of gamma-ray bursts. In our
analysis we use 527 GRB light curves with 64 ms resolution obtained by the
Burst and Transient Source Experiment (BATSE) in the four LAD energy channels,
I--IV: (I) $20-50$ keV, (II) $50-100$ keV, (III) $100-300$ keV, and
(IV) $h\nu>300$ keV. The background is subtracted in each channel using linear
fits to the 1024 ms data.

The method of data analysis is described in \S 2. In \S 3, we study a sample
of the 4 brightest and longest bursts. In \S 4 the average PDS for the full
sample of 527 GRBs is calculated and discussed. In \S 5, we calculate the PDSs
in the separate energy channels and quantify the difference of the temporal
structure between the channels in terms of the PDS slope. We also calculate
the autocorrelation function (ACF) in the separate channels
and compare the results with previous studies of the ACF.
In \S 6, we address dim GRBs and compare them to the
bright ones. 
The results are discussed in \S 7.

%%%%%%%%%%%%%%%%%%%%%%%%%%%%%%%%%%%%%%%%%%%%%%%%%%%%%%%%%%%%%%
\begin{figure*}
\begin{center}
\epsfxsize=19cm
\epsfysize=19cm
\epsfbox{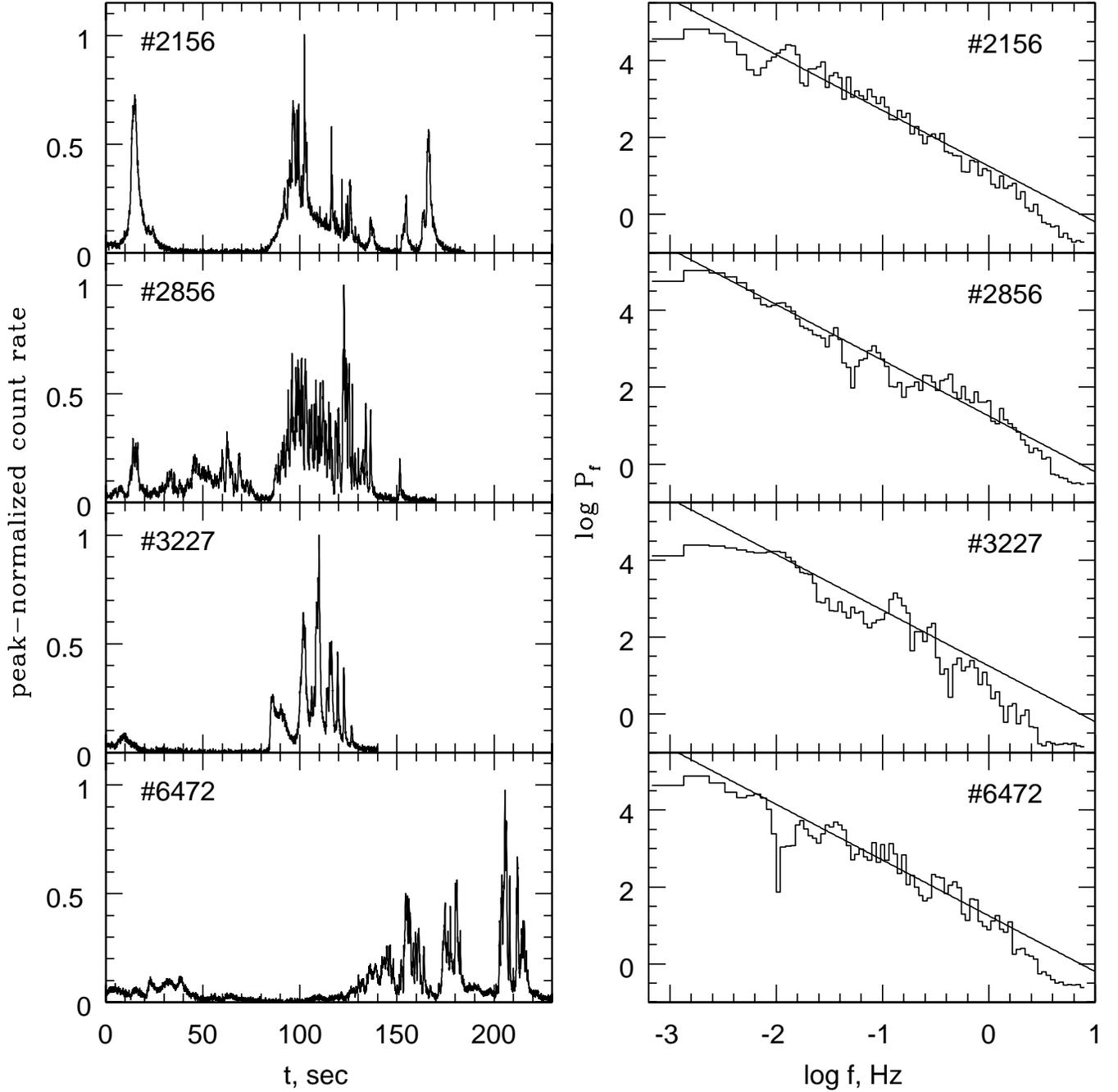}
\end{center}
\caption{
The light curves (in channels II+III) and their PDSs for the four
brightest bursts in the sample with $T_{90}>100$ s. The light curves are taken
peak-normalized and, correspondingly, the PDSs are normalized by $\Cp^2$.
Since the bursts are very bright, their Poisson levels (normalized by $\Cp^2$)
are very low, $P_0\approx 10^{-4}$.
The straight lines show the fit to the average of the 4 PDSs,
$\log P_f=1.25-1.45\log f$ (see Fig.~2).
}
\end{figure*}
%%%%%%%%%%%%%%%%%%%%%%%%%%%%%%%%%%%%%%%%%%%%%%%%%%%%%%%%%%%%%%

%##############################################################################
\section{Data Analysis}

\subsection{The Sample}

Long bursts are of particular
interest since their internal temporal structure can
be studied by spectral analysis over a larger range of time scales. We choose
bursts with durations $T_{90}>20$~s where $T_{90}$ is the time it takes to
accumulate from 5\% to 95\% of the total fluence of a burst summed over
all the four channels, I+II+III+IV.
Hereafter, we measure the brightness of a burst by its peak count rate, $\Cp$,
in channels II+III.
We analyze bursts with $\Cp>100$ counts per time-bin, $\Delta t=0.064$ s.
This condition coupled with the duration condition, $T_{90}>20$~s, gives a
sample of 559 GRBs. The sample still contains GRBs with low fluence, which
are not good for Fourier analysis. We exclude bursts with fluences
$\Phi<32\Cp$. The resulting sample contains 527 bursts.

\subsection{PDS Calculation}

We calculate the Fourier transform, $C_f$, of each light curve, $C(t)$,
using the standard
Fast Fourier Transform method. The power density spectrum, $P_f$, is given by
%the squared Fourier amplitude summed for the two frequencies, $f$ and $-f$,
$P_f=(C_fC_f^*+C_{-f}C_{-f}^*)/2=C_fC_f^*$ since $C(t)$ is real.
The Fourier transform is calculated on a standard grid with a time bin
$\Delta t=64$ ms and a total number of bins $N_{\rm bin}=2^{14}$, which
corresponds to a total time $T\approx 1048$ s since the trigger time.
The light curve of each burst is considered in its
individual time window $(t_1,t_2)$ (see \S 2.4).
In the calculations of the PDS,
we extend the time interval to $(0,T)$ by adding zeros at $(0,t_1)$ and
$(t_2,T)$.
The adding zeros introduces random small-scale fluctuations in the
PDS (see, e.g., Bracewell 1967). We, however, study the PDSs averaged
over adjacent frequencies and/or over a sample of GRBs. Then the 
fluctuations in $P_f$ associated with a specific choice of the grid disappear
and do not affect the results.

\subsection{The Poisson Level}

Poisson noise in the measured count rate affects the light curve on short
time scales, and, correspondingly, affects the PDS at high frequencies.
The Poisson noise has a flat spectrum, $P_f\propto f^0$, introducing
the ``Poisson level'', $P_0$, in a PDS. This level equals the burst total
fluence including the background in the considered time window. The power
spectrum above this level displays the intrinsic variability of the signal
(but see \S 4.2 for the time-window effects). We calculate the
individual Poisson level for each burst and subtract it from the burst PDS.

\subsection{The Time Window}

One would like to see the whole burst in the window. However,
(1) it is difficult to determine exactly the end of a burst because the burst
can always have a weak ``tail'' hidden in the background;
(2) the window should not be very large because inclusion of long weak tails
leads to an increase of the background fluence, $\Phi_b$, without a substantial
increase in the signal fluence, $\Phi$. The resulting low ratio
$\Phi/\Phi_b$ implies a high Poisson level, which
makes the quality of the PDS worse.

To reduce the Poisson level, we cut off the light curves at a time $t_2$
defined so that the signal count rate $C(t)$ does not exceed
$\varepsilon C_{\rm peak}$ at $t>t_2$ and
$C(t_2)=\varepsilon C_{\rm peak}$.
Keeping in mind bursts with a weak beginning, we define the starting window
time, $t_1$, so that $C(t)<\varepsilon C_{\rm peak}$ at
$t<t_1$ and $C(t_1)=\varepsilon C_{\rm peak}$.
In our analysis, $\varepsilon=0.05$ is chosen. We checked that varying
$\varepsilon$ does not significantly affect the results unless
$\varepsilon>0.1$. Note that, for dim bursts, Poisson fluctuations of the
background (which are imprinted on $C(t)$ even after subtraction of the average
background level) may exceed $\varepsilon\Cp$. Then the window $(t_1,t_2)$ is
determined by the background fluctuations rather than by the signal.

%##############################################################################

\section{Individual Bright Bursts}

The brightest and longest bursts are the best ones for Fourier analysis.
In this section, we study the four brightest bursts with $T_{90}>100$ s.
They have trigger numbers \# 2156 (GRB 930201), 2856 (GRB 940302), 3227
(GRB 941008), and 6472 (GRB 941110).

We take the light curves, $C(t)$, summed over channels II and III, in which
the
signal is strongest. To simplify the comparison of different bursts,
we take peak-normalized light curves. Their Fourier transform, $C_f$,
is therefore normalized by $\Cp$, and the PDS, $P_f=C_fC_f^*$, is normalized
by $\Cp^2$.

The four peak-normalized light curves and their PDSs are shown in Figure~1
(we smooth the PDSs on the scale $\Delta \log f = 0.04$ before plotting).
The light curves are very different while the PDSs are similar. They can be
described as a single power law, $\log P_f=A+\alpha\log f$ ($A\approx 1$ and
$\alpha\approx -1.5$) with super-imposed fluctuations $\Delta P_f/P_f\sim 1$.
In spite of the large $\Delta P_f$, the power-law behavior can be seen in
each burst due to the power-law extending over more than two decades in
frequency.

The presence of an underlying power law is an interesting feature of the GRB
temporal behavior. A possible way
to extract it from the noisy individual PDSs is to take
the average PDS over a sample of long GRBs. Then the
fluctuations affecting each individual PDS tend to
cancel each other and one can see the power-law behavior.
The average of the four PDSs is plotted in Figure~2.
One can see that the amplitude of irregularities is reduced after the
averaging and one can measure the slope of the resulting PDS. The
power-law fit in the range $-2.0<\log f<0.2$ gives
$\log P_f = 1.25 - 1.45 \log f$.  
(We use the standard $\chi^2$ fitting
procedure with equal weights for each frequency bin.)

%%%%%%%%%%%%%%%%%%%%%%%%%%%%%%%%%%%%%%%%%%%%%%%%%%%%%%%%%%%%%%
%\medskip
%\begin{figure}[t]
\centerline{\epsfxsize=8.5cm\epsfysize=8.cm {\epsfbox{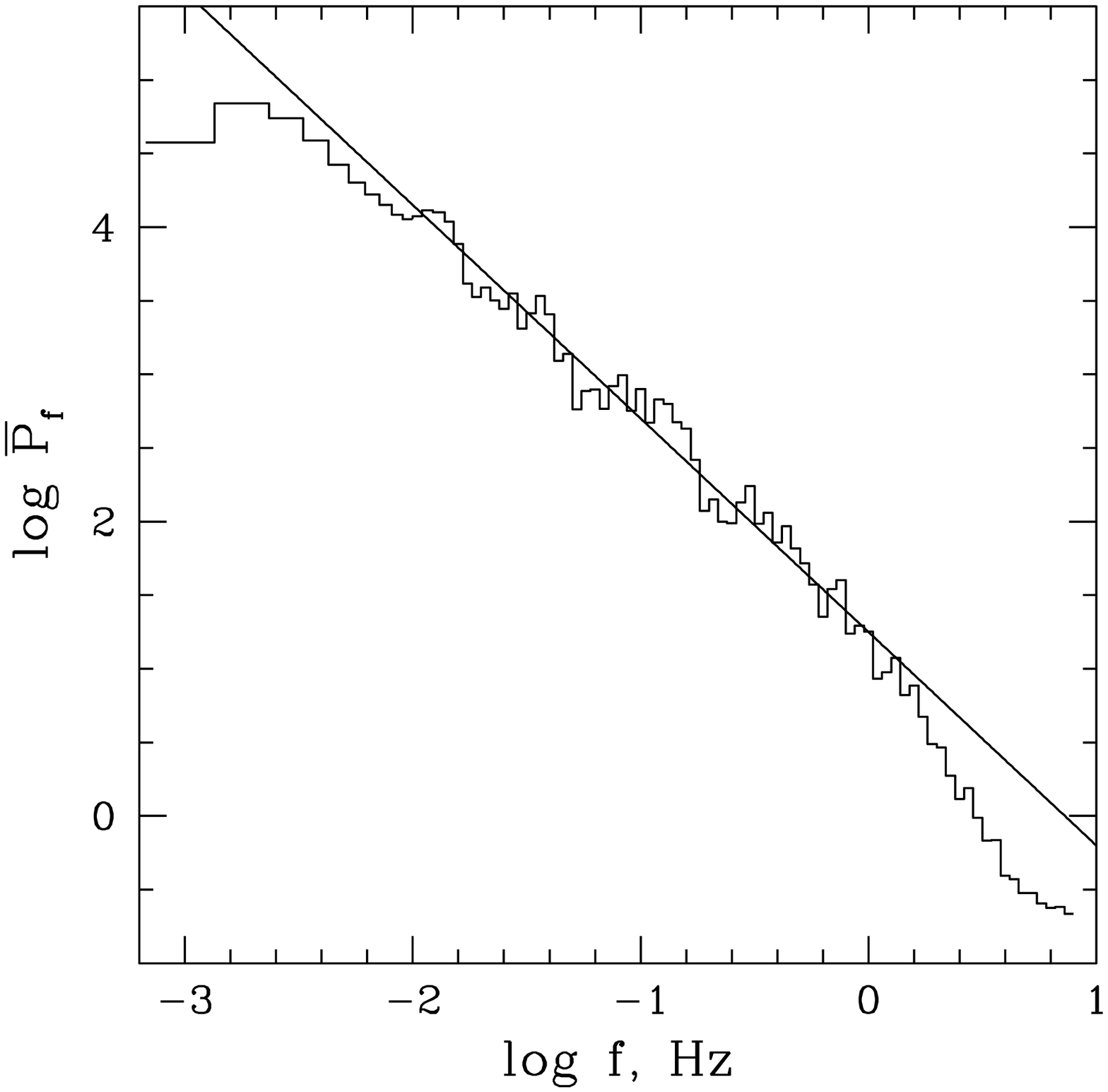}} }
\figcaption{ 
The average of the 4 PDSs shown in Figure~1. The line
shows the power-law fit, $\log P_f=1.25-1.45\log f$, in the
range $-2.0 < \log f < 0.2$.
}
%\end{figure}
\bigskip
%%%%%%%%%%%%%%%%%%%%%%%%%%%%%%%%%%%%%%%%%%%%%%%%%%%%%%%%%%%%%%

The PDS analysis was previously performed for a number of individual bursts.
Belli (1992), for example, analyzed GRBs detected by
the Konus experiment onboard the Soviet probes Venera 11 and Venera 12, and
Giblin et al. (1998) studied individual BATSE bursts.
The power-law PDS can be observed in their longest complex bursts as well.

When turning to GRBs with relatively modest durations, $T_{90}\sim 20 -100$~s,
it becomes more difficult to see the power-law behavior in individual bursts.
The statistical fluctuations, $\Delta P_f/P_f\sim 1$, significantly
affect the apparent PDS slope derived for the available range of frequencies,
which is limited by $\sim T_{90}^{-1}$. However, having a large number of
bursts allows one to study the PDS by averaging over the sample.

%##############################################################################

\section{The Average PDS}

In most of the GRB models (e.g., in the internal shock model), different
bursts
are
produced by one physical mechanism of stochastic nature, i.e., an individual
burst
is a random realization of the same standard process. Stern \& Svensson (1996)
suggested an empirical pulse-avalanche model of this type. They showed that
near-critical avalanches reproduce the statistical characteristics of GRBs
as well as their extreme diversity. With such an approach, individual GRBs are
like pieces of a ``puzzle'', and the features of the standard engine can be
probed
with statistical methods applied to a sufficiently large ensemble of GRBs.

%\newpage

\subsection{The averaging}

The simplest
statistical characteristic of the PDS is the average PDS, $\bar{P}_f$.
The averaging means that we sum up the PDSs of individual bursts with some
weights (normalization) and divide the result by the number of bursts in the
sample. If no normalization is employed, then
the brightest bursts strongly dominate $\bar{P}_f$,
and their individual fluctuations lead to strong fluctuations in $\bar{P}_f$.
The PDSs of relatively weak bursts are suppressed
proportionally to $\Cp^{-2}$ and they are lost in the averaging.
A normalization procedure is thus needed to increase the weight of
relatively weak bursts in the sample. We prefer the peak-normalization
because:

(i) The amplitude of the resulting PDS is practically independent of the burst
brightness (see, e.g., Fig.~1).

(ii) The fluctuations in $\bar{P}_f$ turn out to be minimal and the best
accuracy of the slope is achieved. The distribution of the individual $P_f$
around $\bar{P}_f$ follows a standard exponential law (see Beloborodov et al.
1998), and the amplitude
of fluctuations in $\bar{P}_f$ is given by a simple formula,
$\Delta \bar{P}_f/\bar{P}_f \sim N^{-1/2}$, where $N$ is the number of bursts
in the sample.

The averaging procedure is the following.
(1) For each burst we first determine its individual Poisson level.
(2) We determine the peak, $\Cp$, of the light curve, $C(t)$, with the
background
    subtracted.
(3) We normalize the light curve to its peak.
(4) We calculate the PDS of the normalized light curve.
(5) We normalize the Poisson level by $\Cp^2$ and
    subtract it from the PDS.
(6) We sum up the resulting PDSs over the sample and divide by the number of
bursts. The resulting  average PDS, $\bar{P}_f$, for the 527 peak-normalized
light curves in channels (II+III) is shown in Figure~3.

For comparison, we also plot the average PDS with the
$\sqrt{\Phi}$-normalization, where $\Phi$ is the burst fluence excluding the
background. Each light curve is then normalized by $\sqrt\Phi$, and its PDS
(and
its Poisson level) is normalized by $\Phi$. This normalization gives different
weights (the weights of dim bursts are reduced as compared to the
peak-normalization). The resulting average PDS has a different amplitude.
Nevertheless, $\bar{P}_f$ again follows a power law with approximately the
same slope. This provides evidence that the self-similar behavior with
$\alpha\approx -5/3$ is an intrinsic property of GRBs, rather than an
artifact of the averaging procedure. This interpretation
is also supported by the fact that we observe the power law in the longest
individual bursts. When comparing individual PDSs, $P_f$, with the
peak-normalized
$\bar{P}_f$, one sees that $P_f$ is exponentially distributed around
$\bar{P}_f$,
and the distribution is self-similar with respect to shifts in $f$.

The power-law fit to $\bar{P}_f$ in the range $-1.4 <\log f < -0.1$ gives
$\alpha\approx -1.75$ for the peak-normalized bursts, and $\alpha\approx -1.67$
for $\sqrt{\Phi}$-normalized bursts.
The formal error found from the $\chi^2$ fitting is very small,
$\Delta\alpha\sim \pm 0.01$ (we take the variance,
$\Delta\bar{P}_f/\bar{P}_f=N^{-1/2}$).
Note that the slope in the
peak-normalized case is different from $-1.67$ reported in Beloborodov et al.
(1998). This change is caused by the fact that we have extended the sample to
smaller brightnesses (see \S 6).

The deviation from the power law at the low-frequency end is due to the
finite duration of bursts.
At the high frequency end, there is a break at $\sim 1$~Hz.
The break is observed in the brightest GRBs even without
subtracting the Poisson level (see Beloborodov et al. 1998 and the top
panel in Fig.~8). Note that the break position is the same for the
peak-normalized and $\sqrt{\Phi}$-normalized bursts (see Fig.~3b). It stays
the same
in the separate energy channels (see Fig.~5) and does not depend on the
Poisson
level. One may also see the break in individual long bursts (Figs.~1 and 2).
The break is far too sharp to be explained as an artifact of
the 64 ms time binning, which suppresses the PDS by a factor of
$[\sin(\pi f\Delta t)/(\pi f \Delta t)]^2$ where $\Delta t=64$ ms is the time
bin (cf. van der Klis 1989).

%\medskip
%%%%%%%%%%%%%%%%%%%%%%%%%%%%%%%%%%%%%%%%%%%%%%%%%%%%%%%%%%%%%%
\smallskip
%\begin{figure}
\centerline{\epsfxsize=8.0cm\epsfysize=8cm {\epsfbox{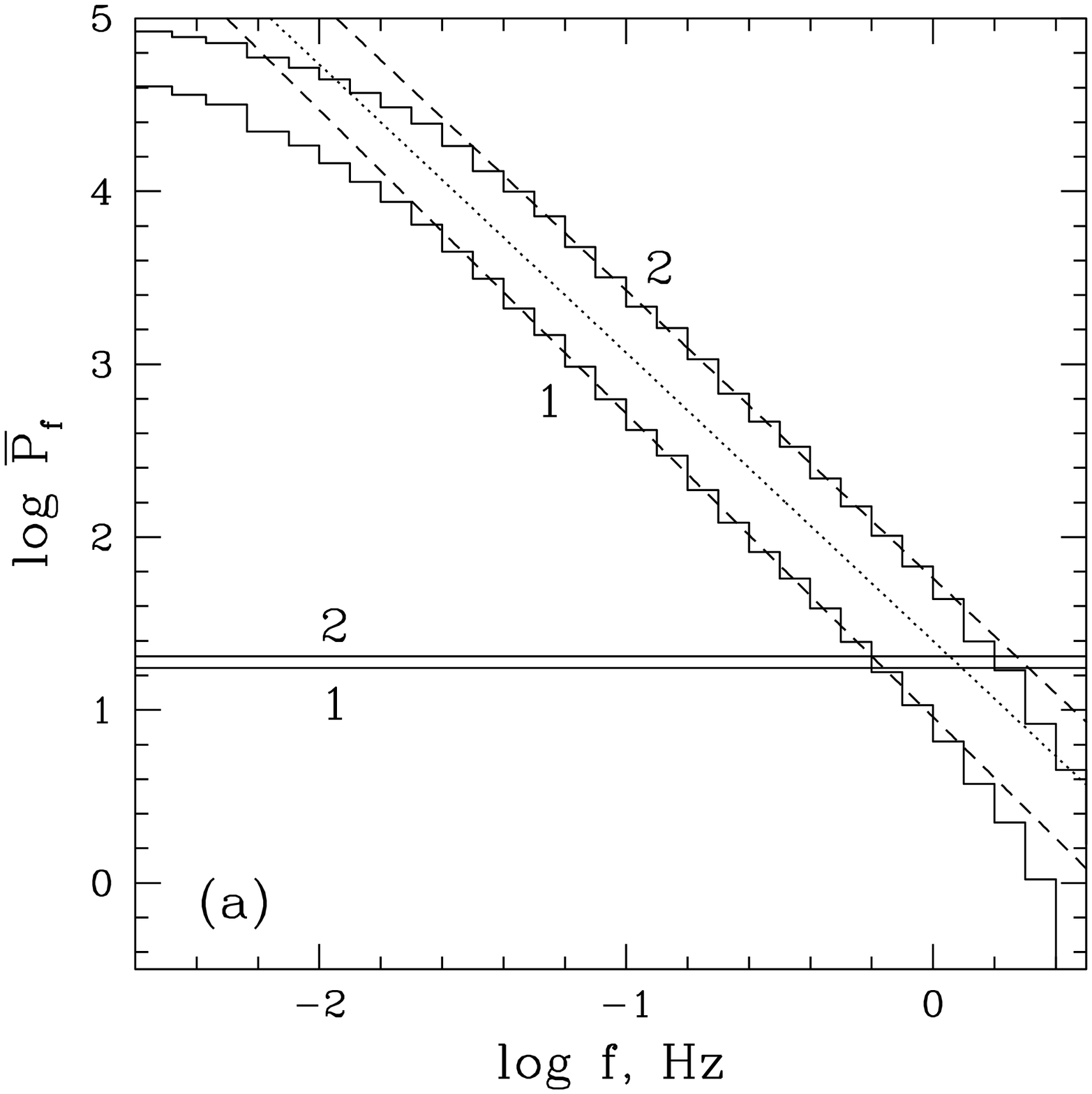}} }
\centerline{\epsfxsize=8.0cm\epsfysize=8cm {\epsfbox{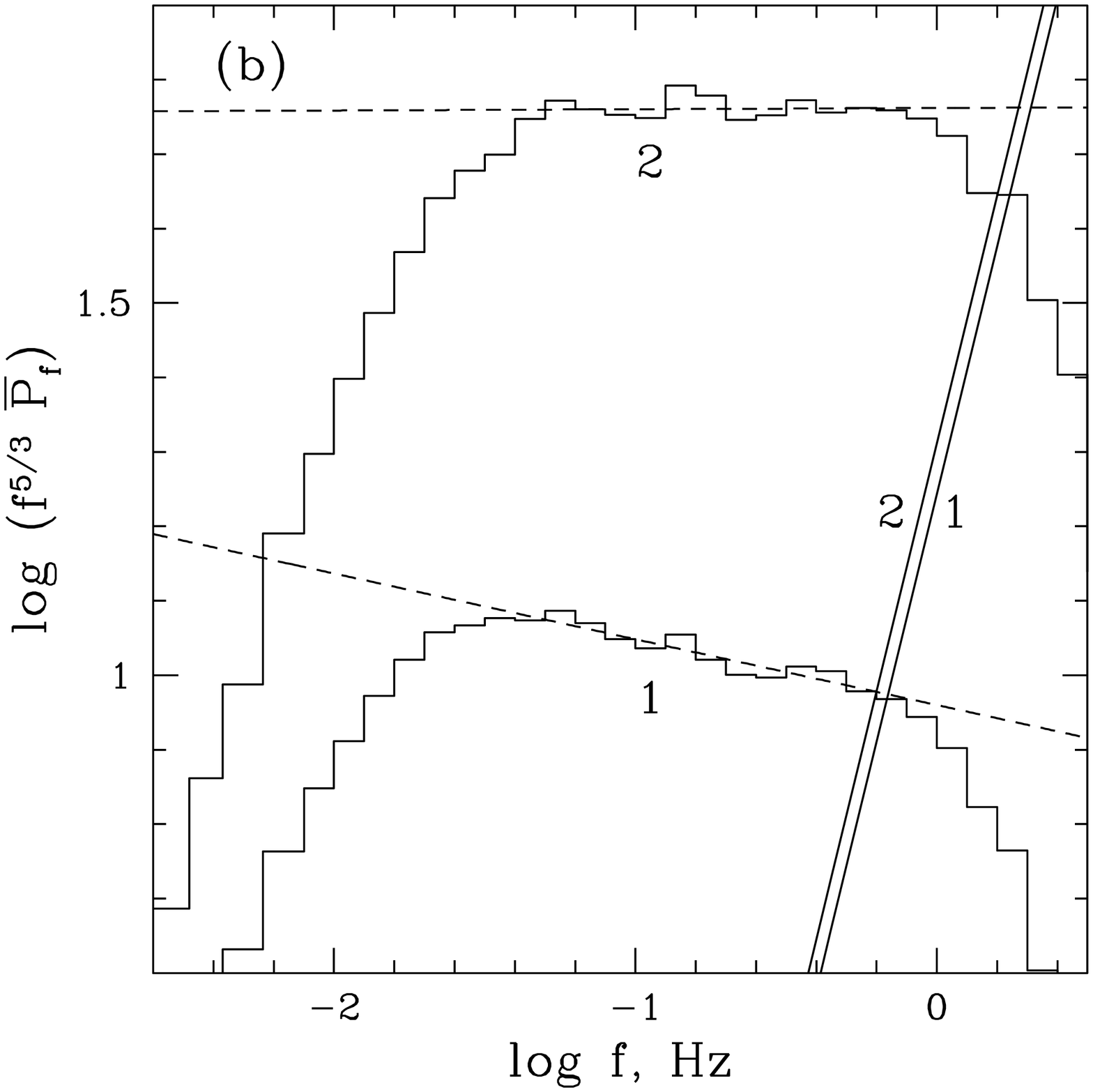}} }
%\bigskip
\figcaption{ 
The average PDS (in channels II+III) for the full sample of
527 GRBs. (a) -- The histograms marked as 1 and 2 correspond to the
peak-normalization and $\sqrt{\Phi}$-normalization, respectively.
The horizontal lines show the corresponding normalized Poisson levels averaged
over the sample. {\it Dashed lines} are the power-law fits in the range
$-1.4<\log f<-0.1$: $\log\bar{P}_f=0.96-1.75\log f$ in the peak-normalized
case and $\log\bar{P}_f=1.76-1.67\log f$ in the $\sqrt{\Phi}$-normalized
case. {\it Dotted line} shows the $-5/3$ slope.
(b) -- Same as (a) except that now $\bar{P}_f$ and the Poisson levels are
plotted multiplied by $f^{5/3}$.
\label{fig3}}
%\end{figure}
%\bigskip
%%%%%%%%%%%%%%%%%%%%%%%%%%%%%%%%%%%%%%%%%%%%%%%%%%%%%%%%%%%%%%

\subsection{The Effects of Finite Signal Duration}

Fourier analysis was designed for physical problems dealing with
linear differential equations. For example, it is usually applied to small
perturbations above a given background solution. The Fourier power spectrum
is also commonly used in the temporal studies of long signals or noises,
e.g., in persistent astrophysical sources. By contrast, GRBs have strongly
non-linear signals with short durations. The typical number of BATSE
time-bins ($\Delta t=64$ ms) in a long GRB is a few $\times 10^3$.
Do the effects of finite duration (i.e., time-window effects)
strongly affect the measured PDS?

%%%%%%%%%%%%%%%%%%%%%%%%%%%%%%%%%%%%%%%%%%%%%%%%%%%%%%%%%%%%%%
\smallskip
%\begin{figure}[t]
\centerline{\epsfxsize=9cm\epsfysize=9cm {\epsfbox{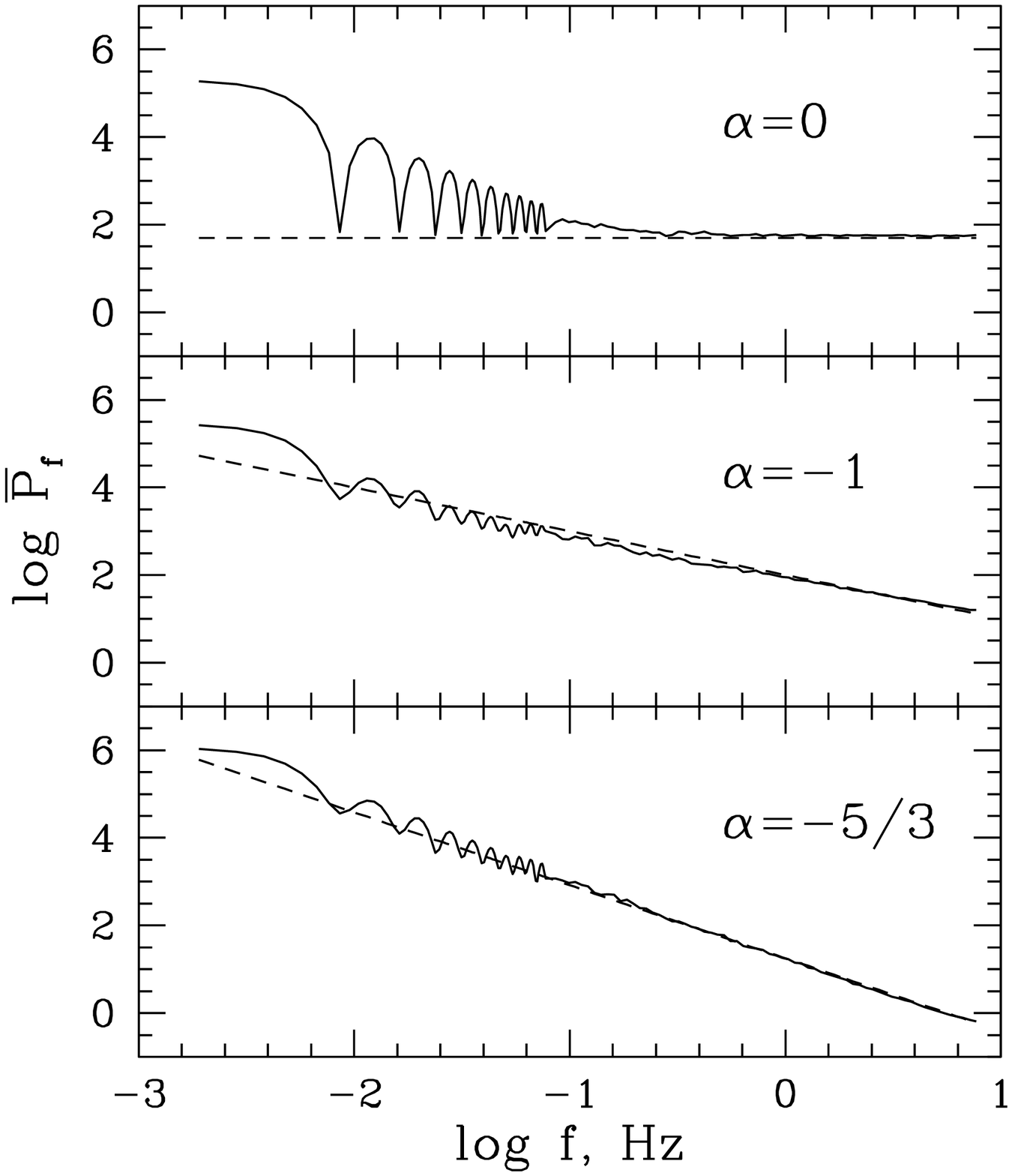}} }
%\bigskip
\figcaption{ 
The average PDS, $\bar{P}_f$, for the 64 fragments of noise
with a power-law PDS of index $\alpha$ (see text for details). {\it Dotted
lines} show the original PDS slope.
The deviation of $\bar{P}_f$ from this slope is due to the time-window effects.
\label{fig4}}
%\end{figure}
\bigskip
%%%%%%%%%%%%%%%%%%%%%%%%%%%%%%%%%%%%%%%%%%%%%%%%%%%%%%%%%%%%%%

The issue is illustrated in Figure~4. We prepared an artificial long
signal with $\bar{P}_f\propto f^{\alpha}$ and exponentially distributed $P_f$,
i.e., the probability to detect $P_f$ at a given $f$ is proportional to
$\exp(-P_f/\bar{P}_f$). The phase structure was taken to be Gaussian (i.e.,
random).
The signal duration is $T_0=2^{17}\Delta t\approx 8400$~s. Then, we cut the
long signal into 64 pieces of equal length $t_0=2^{11}\Delta t\approx 130$ s.
We thus get 64 short signals, each is a random realization/fragment of
the same stationary process characterized by the index $\alpha$.
Analogously to our analysis of real GRBs, we normalize each signal to its peak
and add zeros up to $T=2^{14}\Delta t$ (our standard grid). Then we calculate
the average PDS, $\bar{P}_f$, for the 64 artificial signals. The result is
shown in Figure~4 for the three cases: $\alpha=0$ (Poisson), $\alpha=-1$
(flicker), and $\alpha=-5/3$ (Kolmogorov).

One can see that the time-window effects are strong in the case $\alpha=0$.
The Poisson signal is roughly constant on large time scales. As a result, at
modest $f$, $\bar{P}_f$ is just the power spectrum of the $t_0$-window.
By contrast, in the case $\alpha=-5/3$, we have large-amplitude variations
in the signal on all time scales. The average PDS of the 64 short signals then
reproduces well the intrinsic PDS slope throughout the whole range of
frequencies, down to $f\sim t_0^{-1}$ (see bottom panel in Fig.~4).
Hence, the power law we observe in the average PDS of GRBs
can be interpreted as that GRBs are random short realizations of a
standard process which is characterized by the PDS slope $\alpha\approx -5/3$.
Note, however, that the power spectrum does not provide a complete description
of the signal since the phase structure is not considered.

%#############################################################################

\section{The PDS and the ACF in Channels I, II, III, IV}

What does the average PDS look like in the separate energy channels?
% I, II, III, and IV?
The signal to noise ratio is low in channel I and especially in
channel IV. The number of GRBs for which good PDSs can be obtained in
channels I--IV is therefore limited to the brightest bursts.
We choose a sample of bursts with $\Cp>500$ counts/bin (in channels II+III),
which contains 152 bursts.

The burst analysis is now performed in each channel separately.
We determine the Poisson level of a burst in each channel, $P_0^i$, and
find the peak of the light curve, $\Cp^i$, where
$i=$ I, II, III, IV. Then we perform the peak-normalization:
$C^i(t)\rightarrow C^i(t)/\Cp^i$ and $P_0^i\rightarrow P_0^i/(\Cp^i)^2$.
We set the time window in each channel $(t_1^i,t_2^i)$ as described
in \S 2.4.

%%%%%%%%%%%%%%%%%%%%%%%%%%%%%%%%%%%%%%%%%%%%%%%%%%%%%%%%%%%%%%
\smallskip
%\begin{figure}
\centerline{\epsfxsize=16.5cm\epsfysize=16.5cm {\epsfbox{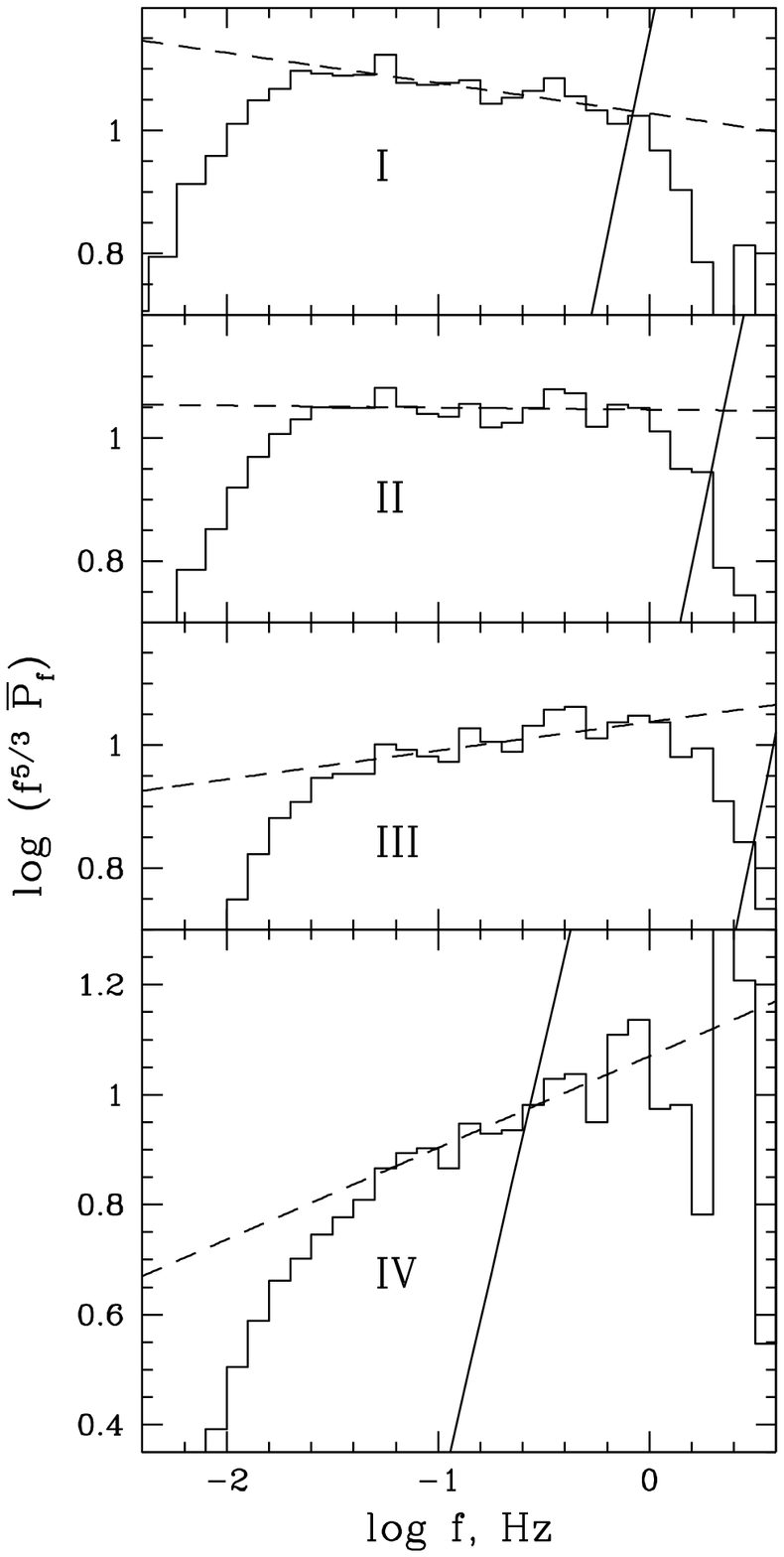}} }
%\bigskip
\figcaption{ 
The average PDS for the sample of 152 GRBs in separate
energy channels, I, II, III, and IV. {\it Dashed lines} show the
power-law fits (see Table~1). {\it Solid lines} show the Poisson level.
\label{fig5}}
%\end{figure}
\medskip
%%%%%%%%%%%%%%%%%%%%%%%%%%%%%%%%%%%%%%%%%%%%%%%%%%%%%%%%%%%%%%

The resulting average PDSs in channels I--IV are shown in Figure~5.
The differences in the slopes are clearly seen.
We fitted the PDSs by power laws, $\log\bar{P}_f=A+\alpha\log f$, in the
range $-1.6 < \log f < 0$. Channel IV is fitted in the range $-1.3 < \log f <
-0.1$. $A$ and $\alpha$ of the  fits are listed in Table~1.

%\medskip
%*************************************************
%\begin{table}
\begin{center}
%{\normalsize
\begin{tabular}{ccccc}
            &   {\bf Table~1.}  &         &        \\ [0.4ex]
\hline\hline
            &                   &         &        \\ [0.4ex]
  Channel   &    I    &   II    &   III   &    IV  \\ [0.8ex]
  $\alpha$  & $-1.72$ & $-1.67$ & $-1.60$ & $-1.50$\\ [0.8ex]
  $A$       &  $1.03$ & $1.05$  &  $1.06$ &  $1.07$\\ [0.8ex]
\hline
\end{tabular}
%}
\end{center}
%\end{table}
%*************************************************
%\medskip

The observed behavior can be briefly described as follows: The ``red'' power
($\bar{P}_f$ at low $f$) decreases at high photon energies and increases at
low
photon energies. It should be compared with the well-known fact that the pulses
in a GRB are narrower in the hard channels (e.g., Norris et al. 1996). The
hardness of emission varies dramatically during a burst and this leads to
different temporal structure in different channels. E.g., pulses observed in
the soft channels may be suppressed in the hard channels. One therefore could
expect that the PDS has different slopes in different channels. Note that,
typically, most of the GRB energy is released in channels II+III, and the
average PDS of {\it bolometric} light-curves approximately follows the $-5/3$
law.

In principle, the autocorrelation function (ACF) contains the same information
as the PDS, since one is the Fourier transform of the other (the
Wiener-Khinchin
theorem). In practice,
the two are not completely equivalent because of the time-window effects and
the presence of a noisy background. On modest time scales, $\simlt 30$~s, the
direct ACF calculation and the calculation via the Fourier transform of the PDS
give the same result to within a few percent (we have calculated the ACF by
both
methods). The average ACF, $\bar{A}(\tau)$, for our sample of 152 bright GRBs
is shown in Figure~6 for each of the four channels. The ACF gets narrower at
high energies, in agreement with previous studies (Fenimore et al. 1995),
except
that our ACF is $\sim 2$ times wider as compared to that obtained by Fenimore
et al. (1995). The ACF width averaged over the channels is in approximate
agreement with that calculated for the bolometric light curves by
Stern \& Svensson (1996).

Note that the average ACF is obtained by summing up the
individual ACFs normalized to unity, i.e., $A(0)=1$. Recalling that the ACF is
the Fourier transform of the PDS, one can see that this normalization is
equivalent
to the normalization of the light curve by $\sqrt{P_{\rm tot}}$ where
$P_{\rm tot}=\int [C(t)]^2 dt = \int P_f df$ is the total power. This
normalization is {\it different} from the peak-normalization we use in the
calculations of the average PDS, i.e., we prescribe different weights to
individual bursts when averaging the PDS and the ACF, respectively.
Therefore, the average
ACF is {\it not} the Fourier transform of the average PDS shown in Figure~5.
This relation holds when the average PDS is calculated with the
$\sqrt{P_{\rm tot}}$-normalization.

The ACF in each channel is perfectly fitted by the stretched exponential:
$\bar{A}(\tau)=\exp(-[\tau/\tau_0]^\beta)$ (see Fig.~7). The parameters
$\tau_0$ and $\beta$ are listed in Table~2. The index $\beta$ is related to
the PDS slope $\alpha$ by the simple relation $\beta\approx -(1+\alpha)$.

%\medskip
%*************************************************
%\begin{table}
\begin{center}
%{\normalsize
\begin{tabular}{ccccc}
            &   {\bf Table~2.}  &         &        \\ [0.4ex]
\hline\hline
            &                   &         &        \\ [0.4ex]
  Channel   &    I    &   II    &   III   &    IV  \\ [0.8ex]
  $\beta$   & $0.73$  & $0.67$  & $0.63$  &  $0.6$ \\ [0.8ex]
  $\tau_0$  & $14.0$  & $10.7$  & $7.3$   &  $5.1$ \\ [0.8ex]
\hline
\end{tabular}
%}
\end{center}
%\end{table}
%*************************************************
%\medskip

The changing $\beta$ demonstrates that the ACF shape changes
from channel to channel, as it should do since the PDS changes.
As a first approximation, the PDS slope is equal to $-5/3$, and the ACF
index $\beta\approx 2/3=5/3-1$.

%%%%%%%%%%%%%%%%%%%%%%%%%%%%%%%%%%%%%%%%%%%%%%%%%%%%%%%%%%%%%%
%\medskip
%\begin{figure}
\centerline{\epsfxsize=9cm\epsfysize=9.2cm {\epsfbox{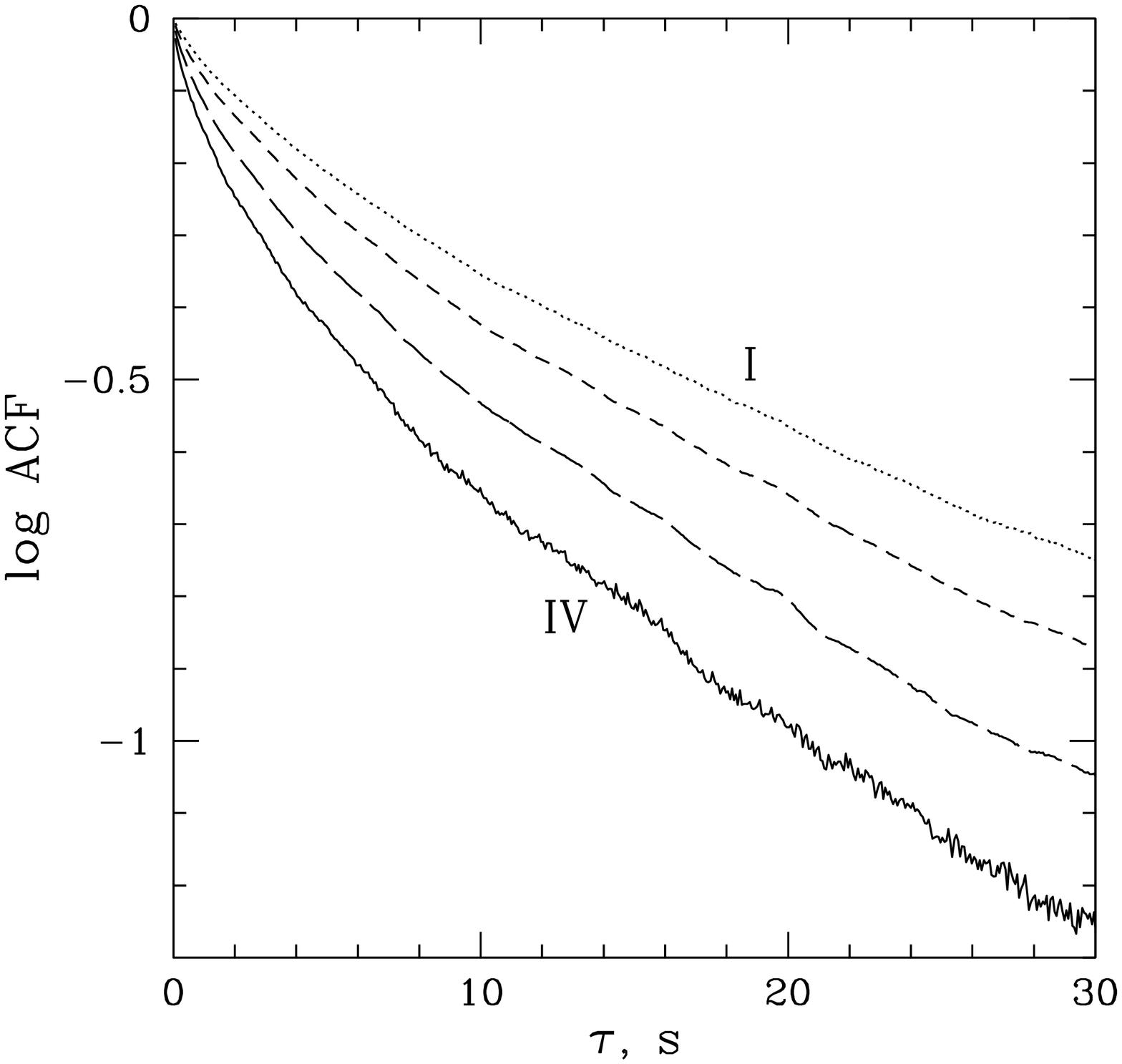}} }
\figcaption{
The average autocorrelation function for the sample of 152 bright
bursts, in channels I--IV.
}
%\end{figure}
%\bigskip
%%%%%%%%%%%%%%%%%%%%%%%%%%%%%%%%%%%%%%%%%%%%%%%%%%%%%%%%%%%%%%

%%%%%%%%%%%%%%%%%%%%%%%%%%%%%%%%%%%%%%%%%%%%%%%%%%%%%%%%%%%%%%
%\medskip
%\begin{figure}
\centerline{\epsfxsize=9cm\epsfysize=9.5cm {\epsfbox{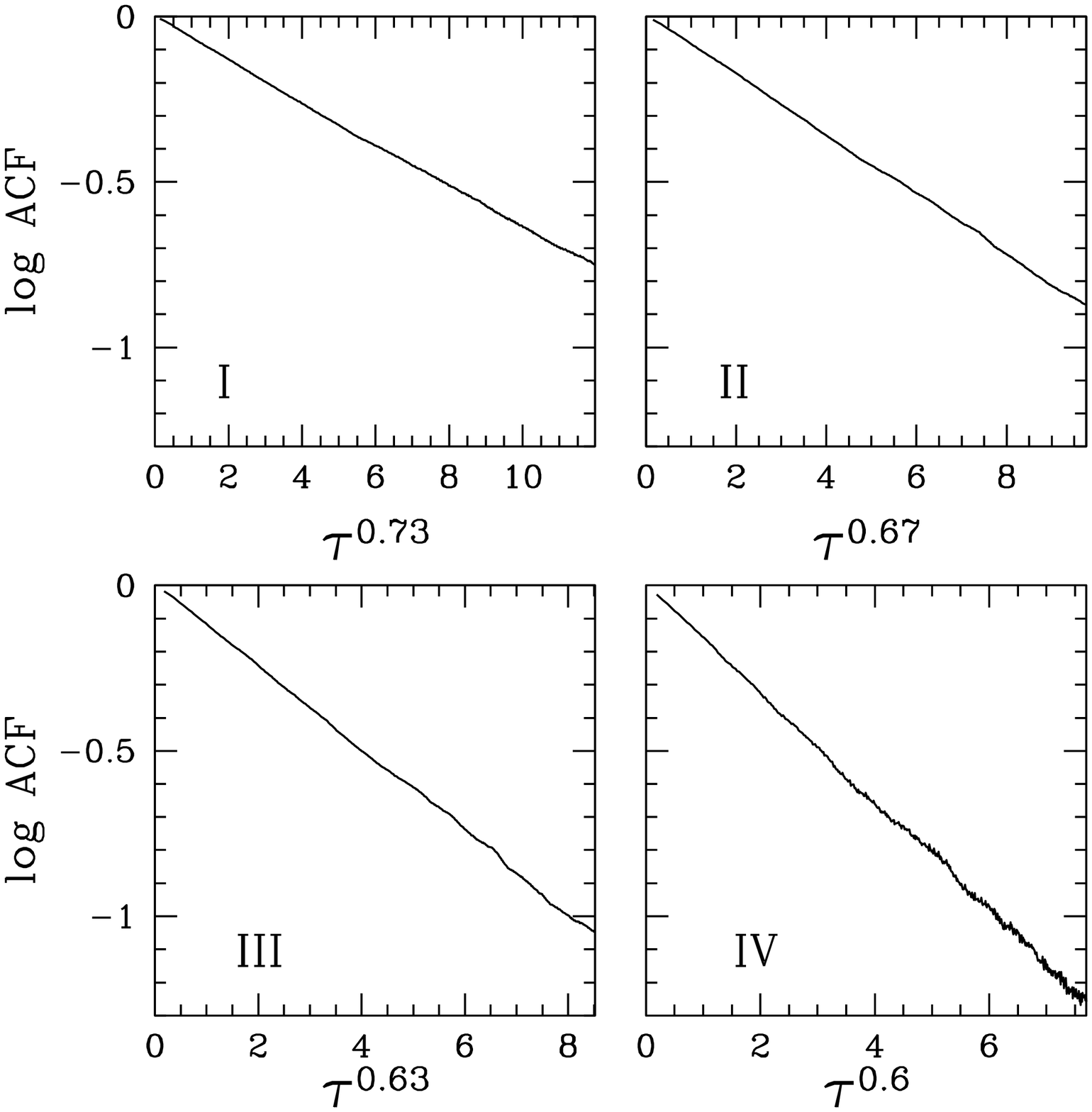}} }
\figcaption{ 
The average ACFs in channels I-IV are plotted against
$\tau^\beta$ which should give a straight line for a stretched exponential.
}
%\end{figure}
\bigskip
%%%%%%%%%%%%%%%%%%%%%%%%%%%%%%%%%%%%%%%%%%%%%%%%%%%%%%%%%%%%%%

The values of $\tau_0$ quantify the ACF width in different channels.
The parameter $\tau_0$ is better defined as compared to measuring
the ACF width at a certain level, e.g., at $e^{-0.5}$ of the maximum level, as
done in Fenimore et al. (1995). Nevertheless, the scaling of $\tau_0$ with
energy is approximately the same, $\tau_0\propto E^{-0.4}$ where $E$ is the
photon energy of the low energy boundary of the channel, see
Fenimore et al. (1995). It should be stressed, however, that the stretching of
$\tau_0$ is {\it not} related to the stretching of any physical time scale of
intrinsic correlations during the burst. The ACF width, $\tau_0$, is
rather related to the position of the {\it breaks} in the PDS, especially
the low frequency break, which in turn is determined by the burst duration.
For an infinitely long signal with a power law PDS, $\tau_0$ would tend to
infinity. This is natural since the power law behavior implies self-similarity,
i.e., the absence of any preferred time scale. Specific time scales are
introduced
only by the breaks in the PDS. We therefore have only two physical
time scales in long GRBs: The first one is associated with the $1$~Hz break and
the second one is associated with the low frequency break due to the finite
burst duration.

%##############################################################################

\section{Dim versus Bright Bursts}

Now we address the following question: Is there any correlation between the
PDS slope and the burst brightness?

\subsection{PDS Slope Correlates with the Burst Brightness}

We take the full sample of 527 light curves in channels II+III and divide
it into 3 groups of different brightnesses: (A) $\Cp>800$, (B) $300<\Cp<800$,
and (C) $100 < \Cp < 300$. Group A contains 91 bursts, group B 222 bursts,
and group C 214 bursts. We have calculated the average PDS for each group
separately employing the peak-normalization. The results are presented in
Figure~8.
We fitted the average PDSs by power laws, $\log P_f= A+\alpha\log f$.  The
parameters $A$ and $\alpha$ of the fits in the range
$-1.5 < \log f < -0.1$ are listed in Table~3.
We conclude that the average PDS of dim bursts gets markedly steeper.

%*************************************************
%\begin{table}
%\caption[ ]{}
\begin{center}
%{\normalsize
\begin{tabular}{cccc}
            & {\bf Table~3.} &         &        \\ [0.4ex]
\hline\hline
            &                &         &        \\ [0.4ex]
  group     &      A         &    B    &   C    \\ [0.8ex]
  $\alpha$  & $-1.63$        & $-1.74$ & $-1.82$\\ [0.8ex]
  $A$       &  $1.04$        &  $0.98$ &  $0.91$\\ [0.8ex]
\hline
\end{tabular}
%}
\end{center}
%\end{table}
%*************************************************
%\medskip

\subsection{Subtraction of the Poisson Level}

In the PDS calculations we subtracted the Poisson level, which is quite high
for dim bursts (see Fig.~8). One therefore should address a technical question:
How well is the intrinsic PDS restored after subtracting the Poisson level?

To investigate this issue we have performed the following test. We take the
sample of 91 bright bursts (group A) and add a strong Poisson
noise (with an average level of 25000 counts/bin) to each burst in the group.
We thus artificially increase the Poisson level by two orders of magnitude.
Alternatively, one may consider this procedure as a rescaling of bright bursts
to a smaller brightness while keeping the Poisson background at the same
level.
Then, we analyze the artificial bursts in the same way as we did with the real
dim GRBs. Note that even after subtraction of
the time-averaged background, its Poisson noise strongly affects the signal and
may create artificial peaks dominating the true peak of the signal. For the
artificial bursts, we know the true peak (which is the peak of the original
bright burst). In real dim bursts we do not know the position of the true peak
and employ the peak search scheme described in Stern, Poutanen, \& Svensson
(1999).

%%%%%%%%%%%%%%%%%%%%%%%%%%%%%%%%%%%%%%%%%%%%%%%%%%%%%%%%%%%%%%
\smallskip
%\begin{figure}
\centerline{\epsfxsize=14.5cm\epsfysize=14.5cm {\epsfbox{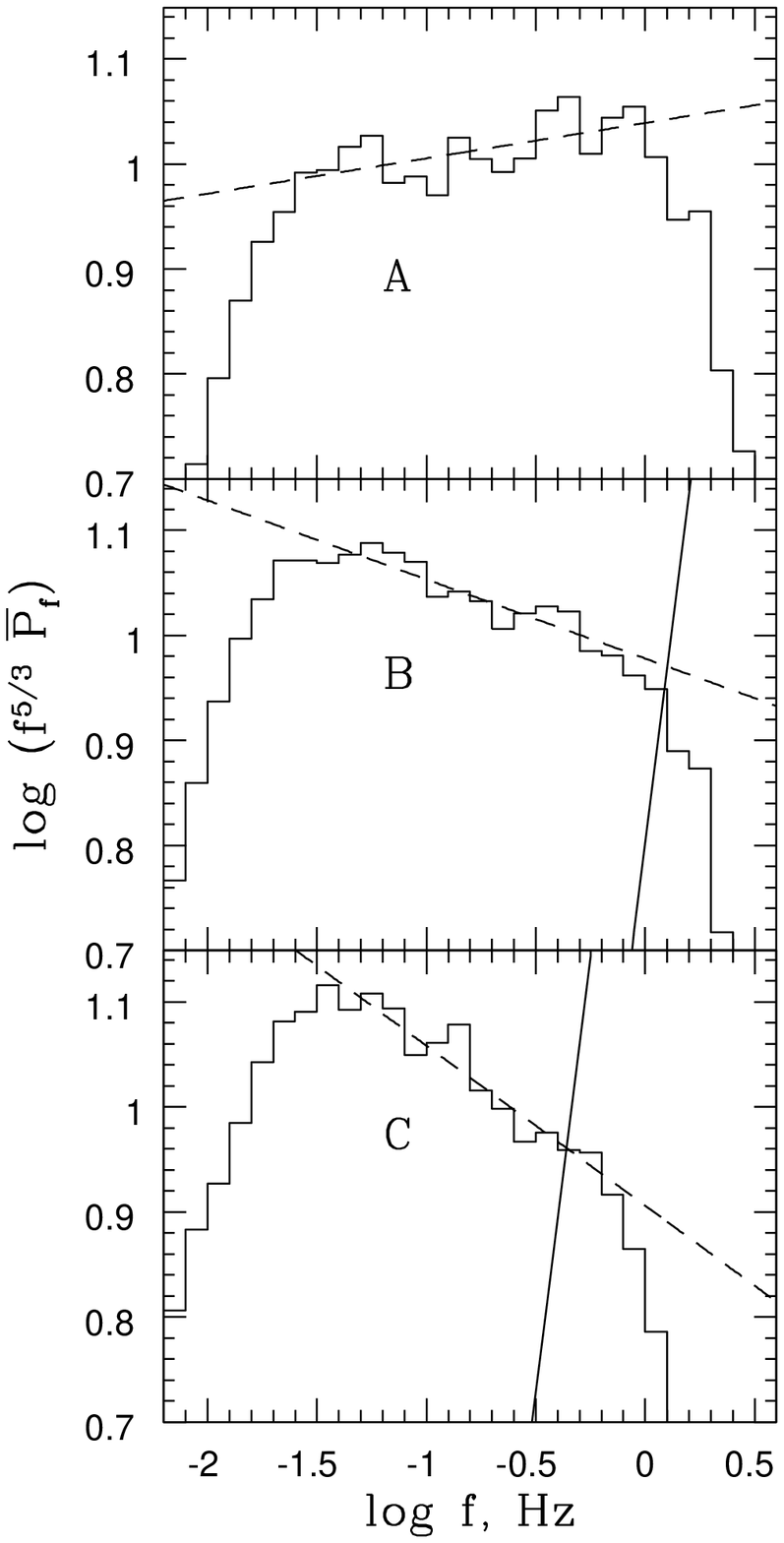}} }
%\bigskip
\figcaption{ 
The average PDSs for the three brightness groups A, B, and C.
{\it Dashed lines} show the power-law fits (see Table~3). {\it Solid lines}
show the Poisson level.
\label{fig8}}
%\end{figure}
\bigskip
%%%%%%%%%%%%%%%%%%%%%%%%%%%%%%%%%%%%%%%%%%%%%%%%%%%%%%%%%%%%%%

The result is compared with the average PDS of the original sample in
Figure~9. We find that the subtraction of the Poisson level allows one to
restore the original PDS, $P_f$, also at frequencies where $P_f$ is well below
the Poisson level. Even the 1 Hz break remains present. We conclude that
the high Poisson level is unlikely to significantly affect the measured PDS
slope.

%##############################################################################

\section{Discussion}

\subsection{Relation to the Average Time Profile}

The average time profile (ATP) of GRBs
was found to follow a stretched exponential of index $1/3$ (Stern 1996).
Is there any relation between the ATP and the average PDS? One should note two
important differences between the ATP and the PDS studies:
(1) the $-5/3$ PDS, though affected by the statistical fluctuations, is
observed in {\it individual} bursts, while the ATP is a purely statistical
property of a large sample of bursts;
(2) the ATP is constructed for GRBs of {\it all durations} (and only in this
case it displays the perfect stretched exponential) while the PDS is studied
for long bursts only. Nevertheless, both the ATP and the PDS characterize the
stochastic process generating GRBs.
Stern (1999) shows that both can be reproduced simultaneously with the
pulse-avalanche model of Stern \& Svensson (1996).

%%%%%%%%%%%%%%%%%%%%%%%%%%%%%%%%%%%%%%%%%%%%%%%%%%%%%%%%%%%%%%
\medskip
%\begin{figure}[t]
\centerline{\epsfxsize=8.0cm\epsfysize=8.0cm {\epsfbox{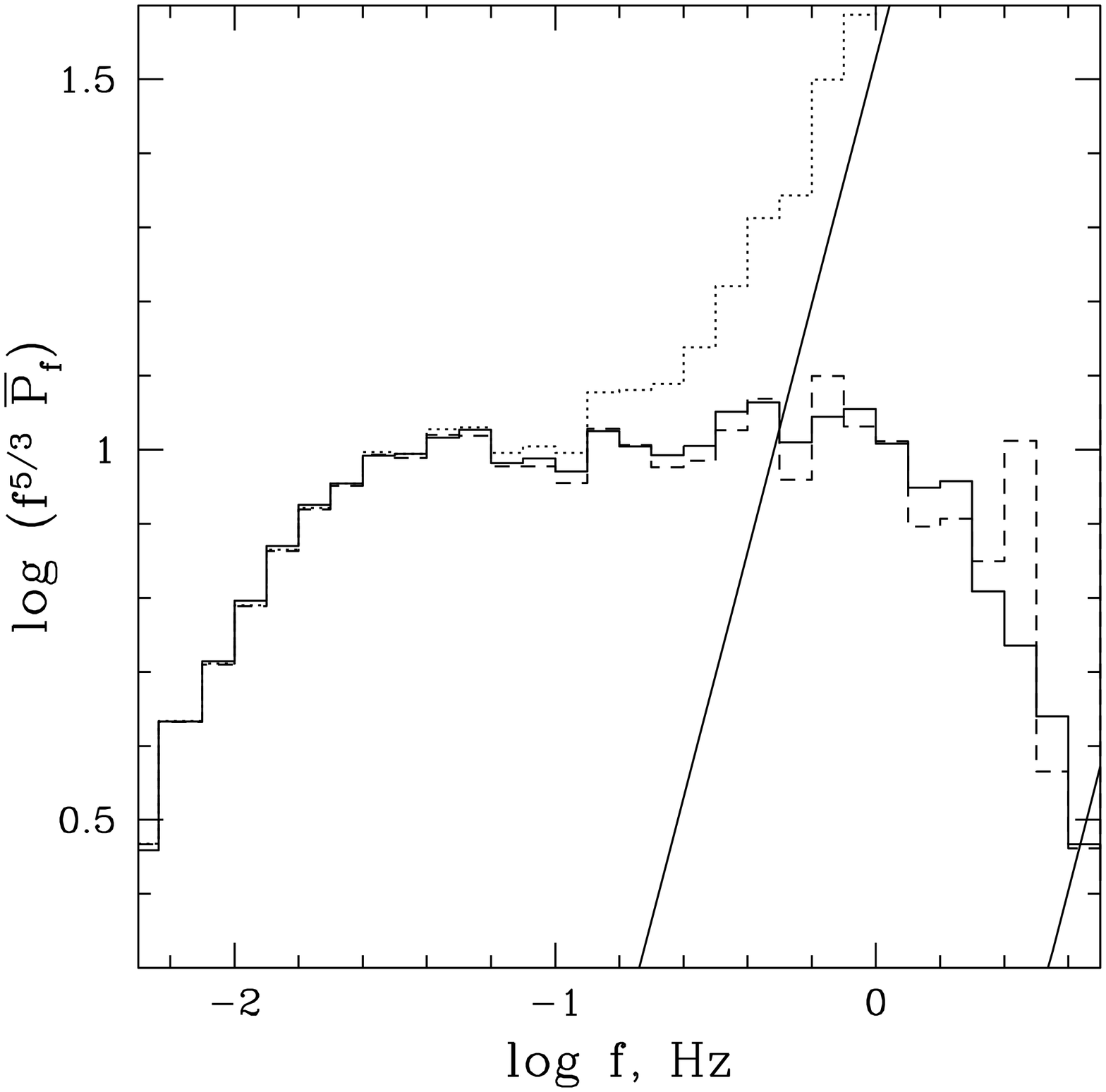}} }
%\bigskip
\figcaption{ 
The average PDS of 91 artificial dim bursts created from
group A of the brightest bursts (see Fig.~8). {\it Solid histogram} -- the
original PDS for group A. {\it Dashed} and {\it dotted histograms}
-- the average PDS of the artificial bursts with and without subtraction of
the Poisson level, respectively. The two solid lines show the average Poisson
levels for the original and the artificial bursts, respectively.
\label{fig8}}
%\end{figure}
\bigskip
%%%%%%%%%%%%%%%%%%%%%%%%%%%%%%%%%%%%%%%%%%%%%%%%%%%%%%%%%%%%%%

\subsection{The 1 Hz Break}

The sharpness of the break in the average PDS appears to be an important
feature that constrains models of GRBs. If the signal is produced in the rest
frame of a relativistic outflow then variations of the outflow Lorentz factor,
$\Gamma$, from burst to burst would smear out the break. The sharpness of the
break then implies a narrow dispersion of $\Gamma$,
$\Delta\Gamma/\Gamma\simlt 2$,
which appears to be unlikely.

Alternatively, the GRB variability may come from the central engine.
The break in the PDS might then be related to a typical time scale, $\sim
1$~s,
in the central engine.

Finally, one may associate the break with the dynamical time scale
corresponding to the inner radius of the optically thin zone of the outflow,
$R_*$. Then the variability on time scales shorter than $t_*=R_*/c\Gamma^2$
can be suppressed. In this case, however, one should explain why $t_*$ stays
the same in different bursts.

\subsection{Dim GRBs}

It has often been hypothesized that dim bursts are at high cosmological
redshifts. For instance, it is necessarily the case if GRBs have approximately
the same intrinsic luminosity, the so-called "standard candle" hypothesis.
We can test this hypothesis using the power spectrum analysis.

Suppose that the dim bursts are intrinsically the same as the bright ones.
Then any difference in their average PDS should be due to a cosmological
redshift.
First consider the bolometric light curves assuming that their average
PDS follows the $-5/3$ power law. It is easy to see that a redshift, $z$, will
not change the PDS slope. As we normalize each bursts to its peak before the
averaging, the effect of a redshift is just stretching the light curve in time,
i.e., precisely the time dilation effect. This will lead to an increase of the
net normalization of the PDS by a factor of $(1+z)^{1/3}$. The slope does not
change since the dilation factor $(1+z)$ is the same for each time scale.

One should, however, recall that we observe
bursts in a limited spectral interval. A redshift then implies a shift of the
signal with respect to our spectral window. E.g., photons detected in
channel III would originally have been emitted in channel IV.
As we know from \S 5, the
PDSs are different in different energy channels. Therefore, one expects that
the PDS slope will change for redshifted GRBs.

We have seen in \S 5 that the PDS of bright bursts flattens in the hard
channels. Hence, a redshift of the bright bursts must be accompanied by a
flattening of the PDS. Contrary to this behavior, we observe that the PDS of
dim bursts {\it steepens}. Hence, the evolution of the PDS with brightness is
inconsistent with the standard candle hypothesis. It implies that the burst
luminosity function is broad and dim bursts are intrinsically weak.

Evidence for a broad luminosity function is also found when looking at the
isotropic luminosities of the bursts with measured redshifts. Note, however,
that the differences in the apparent luminosities could be caused by
orientation effects if GRBs are beamed. One should not therefore exclude that
the total intrinsic luminosities are approximately the same. The different
temporal structure of dim bursts may be an important fact in this respect.
In particular, it suggests that the observed time dilation
of dim bursts (e.g., Norris et al. 1994) may be caused mainly by physical
processes occurring in the bursts rather than by a cosmological redshift.
Note that the intrinsic difference of the temporal
structure of dim GRBs was also found when studying their average time
profile (see Stern et at. 1997).
The rising part of the ATP does not change with decreasing brightness while
the decaying part suffers from time dilation. This behavior is inconsistent
with a cosmological time dilation which should apply equally to both parts
of the ATP.

\subsection{Internal Shocks}

A likely scenario of GRBs involves internal shocks in a relativistic outflow
with a Lorentz factor $\Gamma\sim 10^2$ (see, e.g., Piran 1999 for a review).
The shock develops when an inner faster shell of the outflow tries to overtake
the previous slower shell. The pulses in a burst are then associated with
collisions between the shells. The PDSs predicted by the model were recently
tested against the observed $-5/3$ law (Panaitescu, Spada, \& M\'esz\'aros
1999; Spada, Panaitescu, \& M\'esz\'aros 2000; Beloborodov 1999,2000).
The self-similar temporal structure with $\alpha=-5/3$ can be reproduced by
the model. Further constraints should, however, be imposed by the observed
dependence of $\alpha$ on photon energy. Besides, the phase structure of the
signal, neglected so far, should be taken into consideration
(Beloborodov 2000).

\acknowledgments

This research has made use of data obtained through the HEASARC Online Service
provided by NASA/GSFC.
We thank A.F. Illarionov, I. Panchenko, and J. Poutanen for discussions.
This work was supported by the Swedish Natural Science Research Council,
the Swedish Royal Academy of Science, the Wennergren Foundation for
Scientific Research, a NORDITA Nordic Project grant, the Swedish Institute,
RFBR grant 97-02-16975, and NSF grant PHY94-07194.

%\vspace*{2cm}

\end{document}